\documentclass[sigconf]{acmart}
\usepackage{multirow}
\usepackage{graphicx}
\usepackage{enumitem}
\usepackage{subfigure}  
\usepackage{bbding} 
\usepackage{stfloats}
\usepackage{float}
\usepackage{makecell}
\usepackage{algorithm} 
\usepackage{algpseudocode}
\usepackage{bbding}
\usepackage{balance} 
\usepackage{booktabs}  
\usepackage{multirow}  
\usepackage[table]{xcolor} 
\usepackage{xcolor}

\copyrightyear{2026}
\acmYear{2026}
\setcopyright{cc}
\setcctype{by}
\acmConference[SIGIR '26]{Proceedings of the 49th International ACM SIGIR Conference on Research and Development in Information Retrieval}{July 20--24, 2026}{Melbourne, VIC, Australia}
\acmBooktitle{Proceedings of the 49th International ACM SIGIR Conference on Research and Development in Information Retrieval (SIGIR '26), July 20--24, 2026, Melbourne, VIC, Australia}
\acmDOI{10.1145/3805712.3808510}
\acmISBN{979-8-4007-2599-9/2026/07}
\begin{document}
\title{ReST: A Plug-and-Play Spatially-Constrained Representation Enhancement Framework for Local-Life Recommendation} 

\author{Hao Jiang}
\affiliation{ %
  \institution{Kuaishou Technology}
  \city{Beijing}
  \country{China}}
\email{jianghao11@kuaishou.com}

\author{Long Zhang}
\affiliation{ %
  \institution{Kuaishou Technology}
  \city{Beijing}
  \country{China}}
\email{dragonzhang@mail.ustc.edu.cn}

\author{Guoquan Wang}
\affiliation{ %
  \institution{Kuaishou Technology}
  \city{Beijing}
  \country{China}}
\email{wangguoquan03@kuaishou.com}

\author{Sheng Yu}
\affiliation{ %
  \institution{Kuaishou Technology}
  \city{Beijing}
  \country{China}}
\email{yusheng03@kuaishou.com}

\author{Yang Zeng}
\affiliation{ %
  \institution{Kuaishou Technology}
  \city{Beijing}
  \country{China}}
\email{zhengchengyi@kuaishou.com}

\author{Wencong Zeng}
\affiliation{ %
  \institution{Kuaishou Technology}
  \city{Beijing}
  \country{China}}
\email{zengwencong@kuaishou.com}

\author{Fei Pan}
\affiliation{ %
  \institution{Kuaishou Technology}
  \city{Beijing}
  \country{China}}
\email{panfei05@kuaishou.com}

\author{Peng Jiang}
\affiliation{ %
  \institution{Kuaishou Technology}
  \city{Beijing}
  \country{China}}
\email{jiangpeng@kuaishou.com}

\author{Guorui Zhou$^{*}$}
\affiliation{%
  \institution{Kuaishou Technology}
  \city{Beijing}
  \country{China}}
\email{zhouguorui@kuaishou.com}

\begin{abstract}
Local-life recommendation provides users with convenient access to daily essentials, yet faces two key challenges: 
(1) \textbf{spatial constraints}, where items are exposed only within limited geographic areas, and (2) \textbf{long-tail sparsity}, where popular items dominate user interactions. 
Existing methods predominantly adopt a user-centric view, often failing to address the representation degradation of long-tail items caused by these spatial limitations. 
We argue that an item-centric perspective is better suitable for this domain, as it focuses on enhancing long-tail item representations to align with the spatially-constrained characteristics of local lifestyle services. 
To this end, we propose ReST, a plug-and-play framework for Spatially-Constrained Representation Enhancement. ReST initializes item embeddings via attribute semantics and utilizes a spatially-constrained contrastive learning mechanism. This design adaptively strengthens weak ID representations by aligning them with spatial characteristics, while preserving performance on popular items. Experiments on two real-world datasets and online A/B testing demonstrate that ReST significantly improves performance and exhibits strong plug-and-play flexibility in industrial applications.

\end{abstract}


\vspace{-3.2cm}
\thanks{$^*$Corresponding authors.}
\vspace{-3.2cm}

\maketitle

\begin{figure}[ht]
\vspace{-16pt}
  \includegraphics[width= \linewidth]{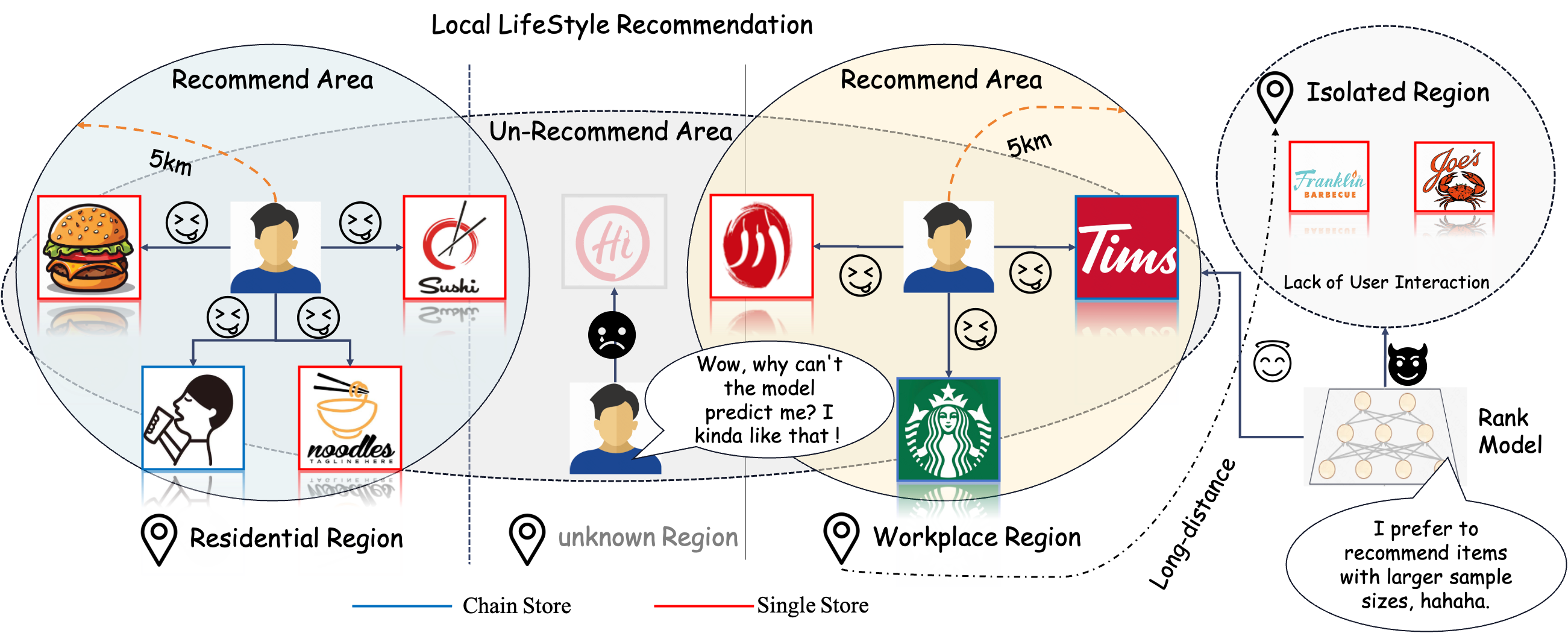}
  \vspace{-20pt}
  \caption{\textbf{Illustration of local-life recommendation: spatial constraints limit accessible candidates, causing inaccurate predictions for disadvantaged items and the Matthew Effect.}}
  \vspace{-20pt}
  \label{f1}
\end{figure}

\section{Introduction}


Online local lifestyle services utilize users’ current geographic locations and historical behaviors to deliver context-aware recommendations, such as nearby shops~\cite{elemedataset, lin2022spatiotemporal}. While increasingly popular on platforms like Kuaishou, Meituan, and Eleme~\cite{chi2023modeling, wang2025fim}, this domain presents unique difficulties compared to traditional scenarios~\cite{schafer2001commerce,celma2010music}, primarily due to \textbf{imbalanced exposure across categories under spatial constraints}.

This problem manifests through two intertwined challenges. First, spatial constraints restrict item visibility ~\cite{zhang2025video,song2025towards,zhang2025enhancing,wang2025episodic,yan2025teleego}. As illustrated in Fig.\ref{f1}, items are typically exposed only within limited geographic ranges, reducing candidate pools and introducing prediction bias in unfamiliar regions~\cite{ofrs, basm}. Second, these constraints exacerbate the long-tail problem. The Matthew effect~\cite{zhang2023empowering} allows popular chains (e.g., McDonald’s) to dominate exposure, while small entities suffer from data sparsity~\cite{luo2023improving,him,lot-crs}. Consequently, capturing latent item relationships becomes difficult, particularly when users rarely access items outside frequent areas~\cite{him}.

Existing methods typically rely on deep user modeling (e.g., GNNs, Transformers), but often fail to overcome spatial limitations~\cite{hidasi2015session, zhang2022enhancing, chen2022building, shin2024attentive, wang2025fim}. Although recent works incorporate spatiotemporal signals via search strategies or architectures like StEN~\cite{lin2022spatiotemporal} and BASM~\cite{basm}, they remain user-centric. These approaches struggle to adapt to dynamic local services and perform poorly under sparse interactions. Furthermore, standard long-tail solutions~\cite{dropoutnet,sigir19,cl4srec} often neglect spatially constrained latent item relationships.

To address these limitations, we propose \textbf{ReST}, a Plug-and-Play \underline{R}epresentation \underline{E}nhancement Framework for \underline{S}patially-Constrained Long-\underline{T}ail Item Recommendation in local lifestyle Services. ReST targets these challenges through two key modules: (1) a \textbf{Meta ID Warm-Up Network} that leverages attribute-level semantics to initialize representations for long-tail items; and (2) a \textbf{Spatially-Constrained ID Representation Enhancement Network  
}, which employs a novel hard sampling strategy to uncover latent item relationships within specific regions. Additionally, a \textbf{dynamic representation alignment strategy} balances collaborative and semantic signals during training. Validated on real-world datasets and the Kuaishou platform, our contributions are as follows:

\begin{itemize}[leftmargin=*]
    \item We design a plug-and-play framework (ReST) that integrates meta ID warm-up and spatially constrained enhancement to robustly model long-tail items.
    \item We propose an item-centric method tailored to the dynamic and sparse nature of local lifestyle services, moving beyond traditional user-centric paradigms.
    \item Offline experiments demonstrate substantial gains, and online A/B testing shows ReST increases \textbf{GMV by 2.804\%} and \textbf{order volume by 1.285\%}.
\end{itemize}

\section{Problem Definition}
Local-life recommendation facilitates daily activities (e.g., dining) by predicting user payments within a fixed radius (e.g., 5 km). Formally, let $\mathcal{U}$ and $\mathcal{V}$ denote user and item sets. A distance-based filter identifies candidates $\mathcal{V}^{D}$ within a limit $D$. We represent user history as $B = (v_1, \ldots, v_{|\mathrm{H}|})$, where $v_i \in \mathcal{V}$ and $|\mathrm{H}|$ is the sequence length. The model recalls spatial items $\mathcal{V}_{u}^{D} \subseteq \mathcal{V}^{D}$ and predicts scores to recommend a top-$k$ list $\{v_{i+1}, \ldots, v_{i+k}\}$ to user $u$.

\section{METHODOLOGY}

\subsection{Model Overview}
As shown in Figure~\ref{f4}(a), the framework comprises two decoupled components: an interchangeable Basic Recommendation Tower and the Spatially-Constrained ID Enhancement Network (SIDENet). SIDENet enhances long-tail representations in local-life scenarios via two sub-modules: the \textbf{Meta ID Warm-up Network}, initializing embeddings via attribute semantics, and \textbf{Spatially-Constrained ID Enhancement}, incorporating hard negative sampling and dynamic alignment.

\subsection{Meta ID Warm-up Network}
This module underpins SIDENet, accelerating ID representation learning by injecting attribute-level semantic information.

\subsubsection{Meta Feature Selection}
Since dense attribute features converge significantly faster than sparse ID embeddings, we leverage them to guide the initialization process. A Squeeze-and-Excitation Network (SeNet)~\cite{hu2019squeezeandexcitationnetworks} is employed to adaptively reweight crucial attributes, such as brand and category:
\begin{equation}
    \mathbf{F}_{i} = \mathbf{SeNet}( \mathbf{E}_{b}, \mathbf{E}_{c}),
\end{equation}
where $\mathbf{F}_{i}$ denotes the reweighted feature representation derived from the item brand $\mathbf{E}_{b}$ and category $\mathbf{E}_{c}$ embeddings.

\subsubsection{Meta Feature Injection}
To integrate these semantic signals into the collaborative latent space, we project $\mathbf{F}_{i}$ using a Multi-Layer Perceptron (MLP) with a residual connection:
\begin{equation}
    \label{eq4}
    \mathbf{E}^{warm}_{i} = \mathcal{MLP}(\mathbf{F}_{i}) + \mathbf{E}_{i},
\end{equation}
where $\mathbf{E}_{i}$ represents the original ID embedding, and $\mathbf{E}^{warm}_{i} \in \mathbb{R}^{d}$ is the enhanced initialization carrying attribute-level semantics.

\subsection{Spatially-Constrained ID Enhancement}
As intrinsic attributes alone may yield indistinguishable representations for similar items, we introduce a self-supervised contrastive learning framework tailored for spatially constrained environments.

\subsubsection{Spatially-Constrained Hard Sampling Strategy} 
\begin{figure*}[tp]
  \vspace{-10pt}
  \includegraphics[width=0.95\linewidth]{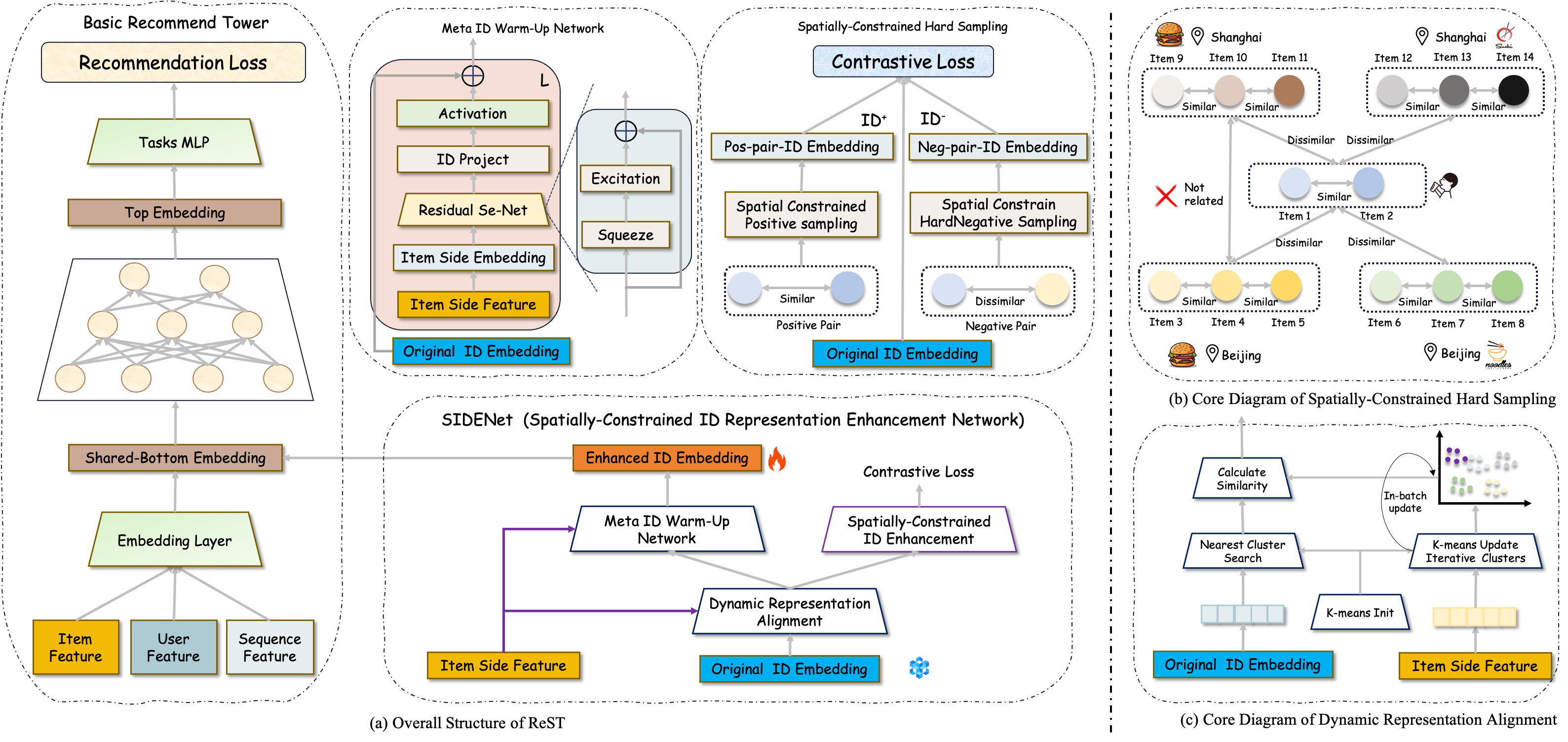}
  \vspace{-10pt}
  \caption{\textbf{The overall architecture of our ReST model. Specifically:(a) shows the architecture of ReST, (b) depicts the core concept of spatially-constrained sampling, and (c) presents the dynamic representation alignment strategy.}}
  \label{f4}
  \vspace{-10pt}
\end{figure*}

Standard co-occurrence sampling~\cite{he2017neural, luo2024qarm} fails in local-life scenarios due to popularity bias and neglecting spatial correlations. We propose a three-stage strategy to filter hard negatives (Figure~\ref{f4}(b)).


\textbf{(1) Prior-Knowledge Hard Sampling.} 
The explicit attributes define initial candidates: items sharing brand or category form positives, while differing ones form negatives. Sampling size $K$ adjusts dynamically based on data distribution to ensure diversity.

\textbf{(2) Similarity-Aware Hard Sampling.} 
To refine the negative pool, we select the most informative samples based on embedding similarity. Given a target embedding $\mathbf{E} \in \mathbb{R}^d$ and a batch of candidate negatives $\mathbf{E}_s \in \mathbb{R}^{n \times d}$, we compute the similarity scores:
\begin{equation}
    \label{eq5}
    \mathrm{s}_i = \operatorname{sim}(\mathbf{E}, \mathbf{E}_s^i), \quad i = 1, 2, \dots, o,
\end{equation}
where $o$ is the batch size. The Top-$K$ items with the highest similarity scores are retained as hard negatives.

\textbf{(3) Spatially-Constrained Negative Sampling.} 
We further filter samples based on spatial proximity, assuming items within a fixed radius share similar interaction patterns. Distances are calculated via the Haversine formula~\cite{robusto1957cosine}:
\begin{equation}
    \label{eq6}
    d_{ij} = \mathbf{Haversine}(\mathbf{GeoHash}_i, \mathbf{GeoHash}_j),
\end{equation}
where $\mathbf{GeoHash}_i$ encodes location ~\cite{liu2014geohash}. Items within a distance threshold are selected as high-quality hard negatives or positives.

\textbf{(4) Final Pairs Construction.} 
We formulate training instances into two categories: (i) \textbf{Positive Pairs}, items sharing attributes (category or brand) within the spatial threshold; and (ii) \textbf{Negative Pairs}, Top-$K$ hard negatives with distinct attributes yet high embedding similarity and spatial proximity.


\subsubsection{Dynamic Representation Alignment Strategy} 
To adaptively balance collaborative signals with attribute-level semantics, we propose a dynamic alignment strategy (Figure~\ref{f4}(c)). This mechanism modulates enhancement strength based on the alignment between item ID embeddings and semantic clusters, stabilizing training and improving robustness. First, item attribute features are projected into a latent semantic space via an MLP:
\begin{equation}
    \label{eq7}
    \mathbf{A} = \mathcal{MLP}(\mathbf{E}_{B} || \mathbf{E}_{c}),
\end{equation}
where $\mathbf{A} \in \mathbb{R}^{n \times d}$. Applying K-means~\cite{arthur2006k} to $\mathbf{A}$ produces cluster centroids $\mathbf{R} \in \mathbb{R}^{|N| \times d}$, where $|N|$ denotes the cluster count. For each item $i$, we identify the nearest centroid index $\mathrm{M} = \textbf{NearestRep}(\mathbf{E}_i, \mathbf{R})$ and compute an adaptive enhancement weight $\alpha_i$:
\begin{equation}
\label{eq10}
\alpha_i = 1 -  \frac{1}{2} \left( \frac{\mathbf{E}_i^\top \mathbf{R}_\mathrm{M}}{\|\mathbf{E}_i\| \cdot \|\mathbf{R}_\mathrm{M}\| + \varepsilon} + 1 \right),
\end{equation}
where $\mathbf{R}_\mathrm{M}$ is the centroid representation and $\varepsilon$ ensures numerical stability. The term in parentheses represents the cosine similarity normalized to $[0, 1]$. A higher similarity implies sufficient semantic alignment, thus reducing the weight $\alpha_i$ to prevent over-regularization. Finally, this weight modulates the injection of semantic information into the warmed-up representation:
\begin{equation}
    \label{eq11}
    \mathbf{E}^{warm}_{i} = \alpha_i \cdot \mathcal{MLP}(\mathbf{S}_{i}) + \mathbf{E}_{i}.
\end{equation}

\subsection{Model Training}
We adopt a multi-task objective combining primary recommendation and auxiliary contrastive tasks for end-to-end optimization. For recommendation, the standard Cross-Entropy loss~\cite{lecun2015deep} is employed:
\begin{equation}
    \label{eq12}
    \mathcal{L}_{rec} = -\frac{1}{|\mathcal{S}|} \sum_{(\boldsymbol{x}, y) \in \mathcal{S}} \left( y \log \mathbf{P}(\boldsymbol{x}) + (1-y) \log (1-\mathbf{P}(\boldsymbol{x})) \right),
\end{equation}
where $\mathcal{S}$ denotes the training set, $|\mathcal{S}|$ is its total size, $x$ represents an input sample, and $y \in [0, 1]$ is the ground-truth label. The output $\mathbf{P}(\boldsymbol{x})$ denotes the predicted probability of $x$ being purchased.

Subsequently, we construct a contrastive learning loss, using \emph{InfoNCE} loss \cite{oord2018representation}, based on the spatially-constrained hard sampling strategy and the dynamic representation alignment strategy. This loss is incorporated as an auxiliary task alongside the primary recommendation objective:
\begin{equation}
    \label{eq13}
    \mathcal{L}_{cl} = -\sum_{i \in \mathcal{O}} \log \left( \frac{\exp(\mathbf{E}_i \cdot \mathbf{E}_{i+}/\tau)}{\sum_{j \in \mathcal{K}} \exp(\mathbf{E}_i \cdot \mathbf{E}_{j}/\tau) + \exp(\mathbf{E}_i \cdot \mathbf{E}_{i+}/\tau)} \right),
\end{equation}
where $\mathcal{O}$ is the item set, $\mathbf{E}_{i+}$ is the positive sample, $\mathcal{K}$ denotes the set of negative samples, and $\tau$ is the temperature parameter. The final objective function integrates both losses:
\begin{equation}
    \label{eq14}
    \mathcal{L}_{total} = \mathcal{L}_{rec} + \alpha_2 \cdot \bar{\alpha} \cdot \mathcal{L}_{cl},
\end{equation}
where $\bar{\alpha}$ is the dynamic enhancement weight derived from Eq.~\eqref{eq10}, and $\alpha_2$ is a hyperparameter balancing the two tasks.

\begin{algorithm}[tbp]
\caption{Dynamic Representation Alignment Strategy}
\label{algo}
\begin{algorithmic}[1]
\Require Item brand $\mathbf{B}$, item category $\mathbf{C}$, item ID embedding $\mathbf{E}_{i}$, number of clusters $|N|$
\Ensure Adaptive enhancement weight $\alpha_i$ for each item
\State Initialize K-means with $|N|$ clusters
\State Initialize parameters of $\mathcal{MLP}(\cdot)$ of Eq.~\eqref{eq7}
\For{$j = 1$ to $MaxIterations$}
    \State $\mathbf{A} \leftarrow \mathcal{MLP}(\mathbf{B}, \mathbf{C})$ \Comment{{Project to semantic space}}
    \State $\mathbf{R} \leftarrow \text{K-means.Update}(\mathbf{A})$ \Comment{{Update cluster centroids}}
    \State $\mathbf{M} \leftarrow \text{K-means.IndexSearch}(\mathbf{E}_i, \mathbf{R})$ \Comment{{Find nearest centroid}}
    \State Compute $\alpha_i$ using Eq.~\eqref{eq10} \Comment{{Enhancement weight}}
\EndFor
\State \Return $\alpha_i$
\end{algorithmic}
\end{algorithm}
\vspace{-4pt}

\vspace{2pt}
\section{Experiment}

\subsection{Experimental Settings}

\noindent\textbf{Datasets.}
As shown in Table \ref{tabcom}, we conduct experiments on two large-scale industrial datasets:
(1) \textbf{Eleme}~\cite{elemedataset}: A large-scale takeaway benchmark incorporating rich spatial-temporal features.
(2) \textbf{Kuaishou}: Sampled from one year of paid transactions on Kuaishou's local lifestyle platform.


\begin{table}[ht]
\vspace{-8pt}
\centering
\caption{Dataset Statistics.}
\vspace{-10pt}
\begin{tabular}{c|cccccc}
\toprule
Datasets & Sample & User & Item  & Avg. action length \\
\hline
Eleme & 1,440,600 & 70,389 & 2,092,944 & 41.008 \\
Kuaishou & 1,046,100 & 72,805 & 81,487  & 8.895 \\
\bottomrule
\end{tabular}
\label{tabcom}
\vspace{-10pt}
\end{table}

\noindent\textbf{Baselines.}
We compare ReST against two categories:
\textit{Representation Learning Methods} (DIN~\cite{din}, RCL~\cite{rcl}), focusing on latent user-item interactions; and \textit{Spatially-Constrained Methods} (TRISAN~\cite{trisan}, OFRS~\cite{ofrs}, BASM~\cite{basm}), which model spatiotemporal dynamics.


\noindent\textbf{Evaluation Metrics.} 
We adopt three standard metrics to assess performance: AUC~\cite{auc}, {MRR}~\cite{mrr} and {NDCG@$k$}, $k\in\{5,10\}$.

\begin{table*}[t]
\centering
\caption{Overall performance on All (full dataset) and Cold-start settings (items appearing < 3 times).}
\label{tab:overall_normal_cold}
\vspace{-8pt}

\small  
\renewcommand{\arraystretch}{1.15} 
\setlength{\tabcolsep}{4.96pt} 
\definecolor{HighlightGray}{gray}{0.92}

\begin{tabular}{l ccc ccc ccc ccc}
\toprule
\multirow{3}{*}{\textbf{Methods}} 
& \multicolumn{6}{c}{\textbf{Kuaishou}} 
& \multicolumn{6}{c}{\textbf{Eleme}} \\
\cmidrule(lr){2-7} \cmidrule(lr){8-13}

& \multicolumn{3}{c}{All} & \multicolumn{3}{c}{Cold-start}
& \multicolumn{3}{c}{All} & \multicolumn{3}{c}{Cold-start} \\
\cmidrule(lr){2-4} \cmidrule(lr){5-7} \cmidrule(lr){8-10} \cmidrule(lr){11-13}

& AUC & MRR & \footnotesize{NDCG@5} & AUC & MRR & \footnotesize{NDCG@5}
& AUC & MRR & \footnotesize{NDCG@5} & AUC & MRR & \footnotesize{NDCG@5} \\
\midrule

DIN \cite{din}
& 0.6946 & 0.3977 & 0.3911
& 0.6720 & 0.3846 & 0.3775
& 0.5653 & 0.1911 & 0.1615
& 0.5597 & 0.2184 & 0.1967 \\

TRISAN \cite{trisan}
& 0.7039 & 0.3968 & 0.3978
& \underline{0.6741} & 0.3790 & 0.3779
& 0.5283 & 0.2015 & 0.1753
& 0.5237 & 0.1997 & 0.1711 \\

OFRS \cite{ofrs}
& 0.7216 & 0.4074 & 0.4101
& 0.5409 & 0.2038 & 0.1797
& 0.5220 & 0.1920 & 0.1634
& 0.5067 & 0.1836 & 0.1526 \\

BASM \cite{basm}
& 0.7193 & 0.4168 & 0.4187
& 0.6623 & \textbf{0.3974} & \underline{0.3953}
& \underline{0.5752} & \textbf{0.2318} & \textbf{0.2115}
& 0.5714 & \textbf{0.2263} & \textbf{0.2049} \\

RCL \cite{rcl}
& 0.7078 & \textbf{0.4327} & \underline{0.4332}
& 0.6397 & 0.3186 & 0.2942
& 0.5730 & 0.1542 & 0.1413
& \underline{0.5736} & 0.1523 & 0.1399 \\

\rowcolor{HighlightGray}
\textbf{ReST (Ours)}
& \textbf{0.7436} & \underline{0.4266} & \textbf{0.4345}
& \textbf{0.6866} & \underline{0.3871} & \textbf{0.3954}
& \textbf{0.5830} & \underline{0.2194} & \underline{0.1973}
& \textbf{0.5824} & \underline{0.2223} & \underline{0.2018} \\
\bottomrule
\end{tabular}
\vspace{-6pt}
\end{table*}

\begin{table}[t]
\centering
\small
\definecolor{DarkRed}{RGB}{160, 0, 0}
\newcommand{\gain}[1]{\textcolor{DarkRed}{\textbf{#1}}}
\caption{Performance gains of ReST across diverse methods.}
\vspace{-8pt}
\label{tab2}
\renewcommand{\arraystretch}{1.06}
\setlength{\tabcolsep}{4pt} 

\begin{tabular}{l cccc}
\toprule
\textbf{Method} & \textbf{AUC} & \textbf{MRR} & \textbf{NDCG@5} & \textbf{NDCG@10} \\
\midrule
DIN         & 0.6946 & 0.3977 & 0.3911 & 0.4595 \\
DIN+ReST    & 0.7059 & 0.4061 & 0.4089 & 0.4778 \\
\textit{Gain} & \gain{+1.62\%} & \gain{+2.12\%} & \gain{+4.55\%} & \gain{+3.98\%} \\
\midrule 

TRISAN      & 0.7039 & 0.3968 & 0.3978 & 0.4633 \\
TRISAN+ReST & 0.7454 & 0.4426 & 0.4478 & 0.5125 \\
\textit{Gain} & \gain{+5.90\%} & \gain{+11.54\%} & \gain{+12.57\%} & \gain{+10.62\%} \\
\midrule

BASM        & 0.7193 & 0.4168 & 0.4187 & 0.4796 \\
BASM+ReST   & 0.7201 & 0.4227 & 0.4204 & 0.4852 \\
\textit{Gain} & \gain{+0.11\%} & \gain{+1.41\%} & \gain{+0.41\%} & \gain{+1.17\%} \\
\bottomrule
\end{tabular}
\vspace{-12pt}
\end{table}

\subsection{Overall Performance}
\textbf{Performance on Full Dataset.} 
As shown in Table~\ref{tab:overall_normal_cold}, ReST secures the highest AUC scores across both datasets. On Kuaishou, ReST outperforms strong baselines with an AUC of 0.7436. Even on the denser Eleme, where spatiotemporal methods like BASM~\cite{basm} remain competitive (excelling in MRR), ReST leads in AUC (0.5830). This indicates that ReST balances ranking quality and discriminative capability, capturing preferences in complex distributions.

\textbf{Robustness in Cold-start.} 
ReST remains robust in the cold-start settings  (items with $<3$ interactions). Unlike methods such as OFRS suffering drastic degradation from sparsity (e.g., Kuaishou AUC dropping 0.7216 to 0.5409), ReST maintains high accuracy. Furthermore, ReST surpasses robust baselines like BASM in AUC across both datasets. This confirms our framework mitigates cold-start limitations inherent to traditional ID-based embeddings.

\textbf{Generalization Ability Analysis.} 
\label{exp1}
Table~\ref{tab2} shows ReST consistently enhances diverse backbones, verifying its plug-and-play generalization.
The largest gains occur in TRISAN~\cite{trisan} (AUC +5.90\%), indicating our \textit{SIDENet} module effectively supplements missing geographic constraints and alleviates sequential sparsity.
Similarly, representation-based models like DIN achieve notable uplifts (e.g., +1.62\% AUC, +4.55\% NDCG@5). 
Even for advanced baselines like BASM~\cite{basm}, ReST yields consistent improvements (+1.41\% MRR).


\begin{figure}[htp]
  \includegraphics[width= \linewidth]{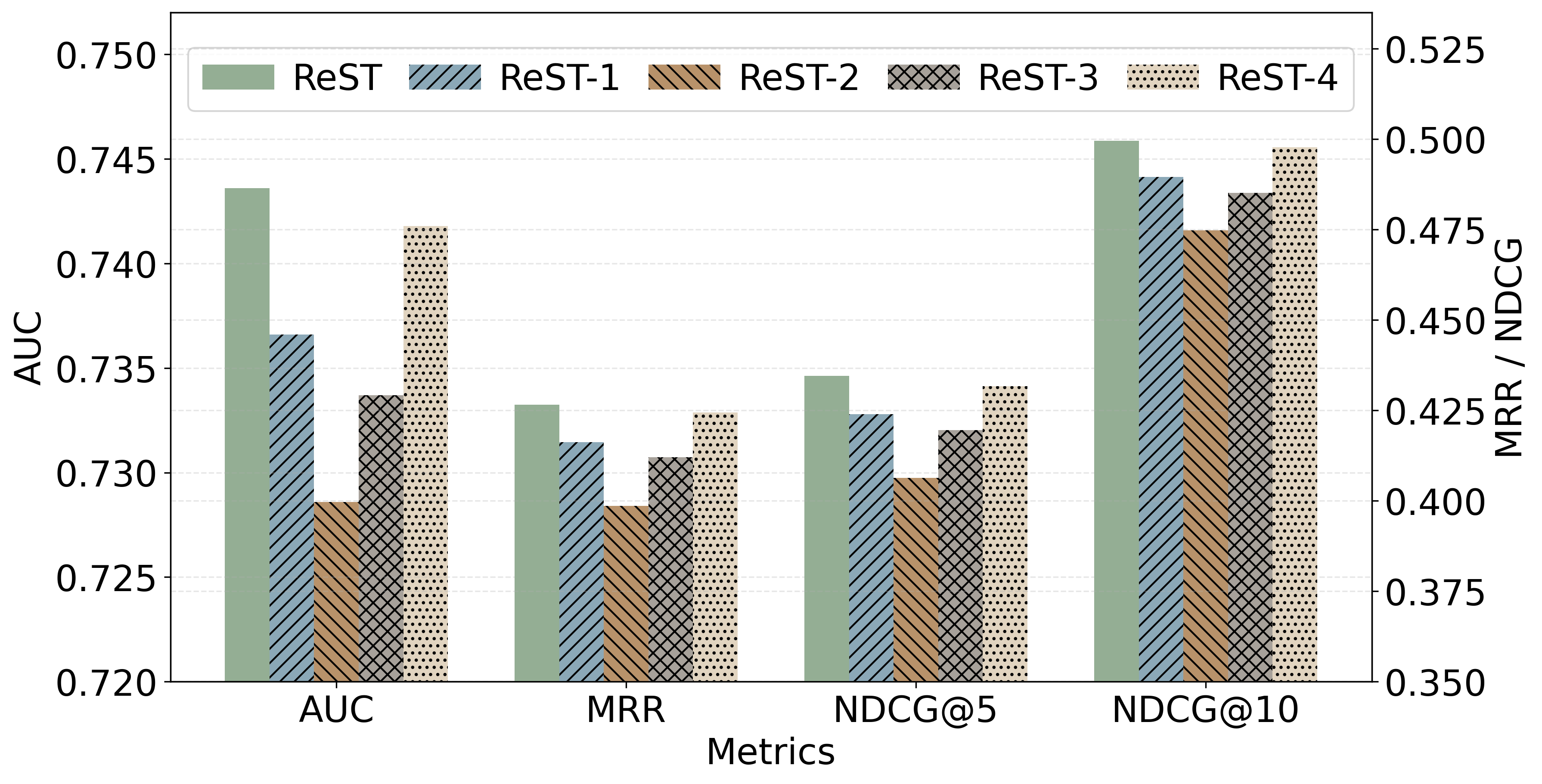}
  \vspace{-22pt}
  \caption{\textbf{Ablation study on ReST.}}
  \label{figab}
  \vspace{-9pt}
\end{figure}

\begin{table}
\caption{Performance comparison with different distance thresholds (km). ``$\infty$'' denotes no distance constraints are applied for contrastive samples.}
\vspace{-6pt}
\begin{tabular}{cc|cccc}
\toprule
Pos & Neg  & AUC & MRR & NDCG@5 & NDCG@10 \\
\hline
5 & 5 & 0.7329 & 0.4052 & 0.4132 & 0.4810 \\
5 & 10 & 0.7337 & 0.4082 & 0.4146 & 0.4826 \\
5 & 30 & 0.7275 & 0.3976 & 0.4035 & 0.4738 \\
10 & 5 & 0.7340 & 0.4094 & 0.4168 & 0.4842 \\
10 & 10 & 0.7337 & 0.4089 & 0.4161 & 0.4836 \\
10 & 30 & 0.7308 & 0.4040 & 0.4109 & 0.4792 \\
30 & 5 & 0.7270 & 0.4054 & 0.4114 & 0.4777 \\
30 & 10 & \textbf{0.7436} & \textbf{0.4266} & \textbf{0.4345} & \textbf{0.4996} \\
30 & 30 & 0.7343 & 0.4137 & 0.4204 & 0.4869 \\
$\infty$ & $\infty$ & 0.6811 & 0.3310 & 0.3260 & 0.4077 \\
\bottomrule
\end{tabular}
\label{hy1}
\vspace{-9pt}
\end{table}

\subsection{Ablation Study}
\subsubsection{\bf{Main Components}}
We evaluate four variants to isolate component contributions: 
(1) \textbf{ReST-1} replaces hard negative sampling with random in-batch sampling;
(2) \textbf{ReST-2} removes brand constraints;
(3) \textbf{ReST-3} removes category constraints;
(4) \textbf{ReST-4} discards the warm-up network.

As shown in Figure \ref{figab}, all variants show performance degradation compare to the complete ReST model, demonstrating the effectiveness of each component. Specifically, ReST-1 experiences a notable decline in AUC by 0.70\%, indicating that hard negative sampling is crucial for learning discriminative representations. ReST-2 and ReST-3 show more significant performance drops, with AUC decreasing by 1.49\% and 1.99\% respectively, which validates that both brand and category features are essential for effective positive sample selection. ReST-4 shows the smallest performance drop (0.17\% in AUC), suggesting that while the warm-up network contributes to the model's performance, the core representation enhancement mechanisms remain effective even without it.

\begin{figure*}[t]
  \includegraphics[width= \linewidth]{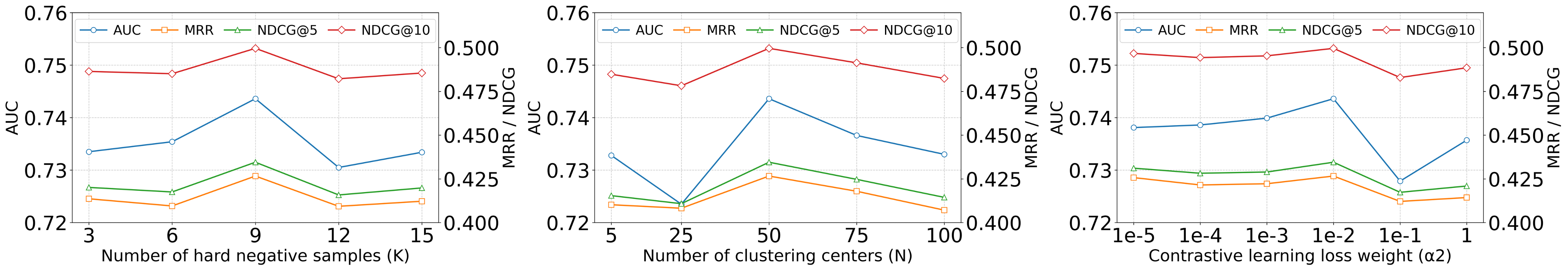}
  \vspace{-18pt}
  \caption{\textbf{Performance comparison with different numbers of hard negative samples.}}
  \vspace{-9pt}
  \label{hy2}
\end{figure*}

\begin{figure}[t]
  \includegraphics[width= \linewidth]{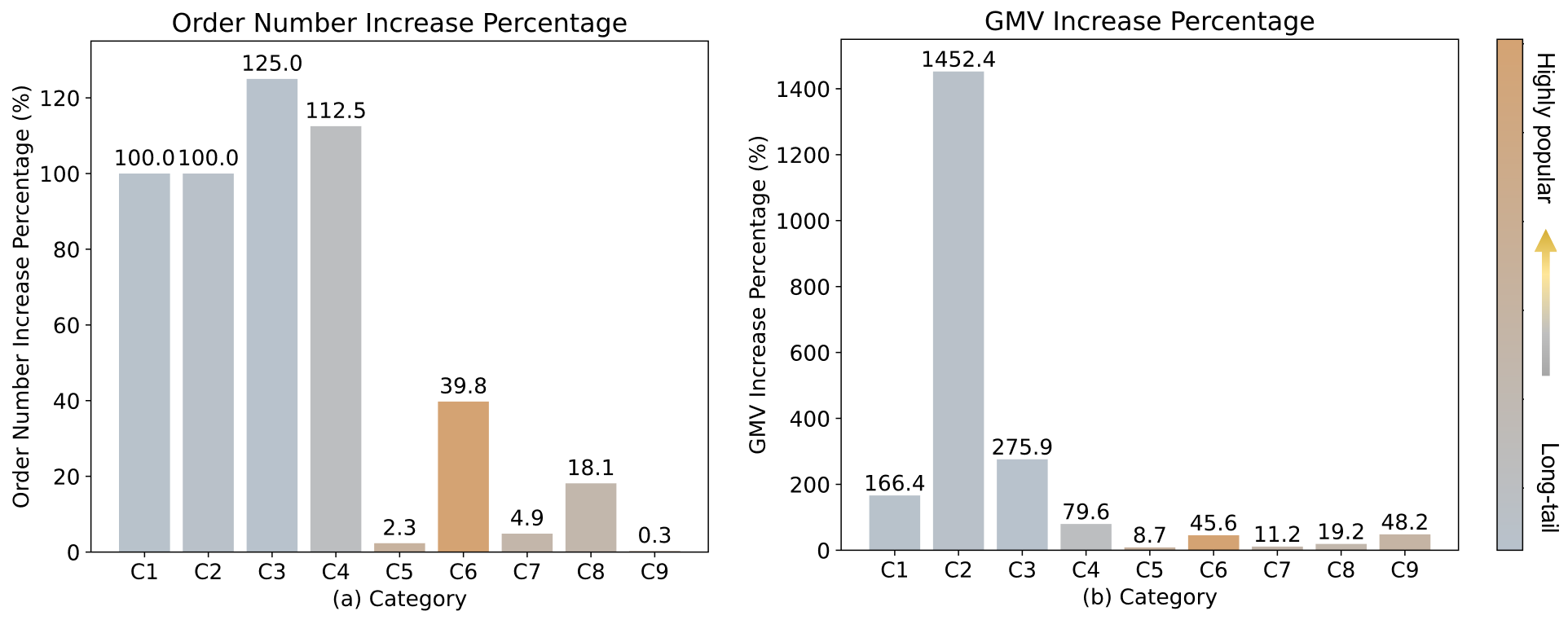}
  \vspace{-16pt}
  \caption{Orders and GMV gains across diverse popularity. }
  \label{vis1}
  \vspace{-16pt}
\end{figure}

\subsubsection{\bf{Effect of Hyper-parameters}}
We investigate the impact of several hyperparameters on model performance: (1) distance thresholds for positive and negative samples, (2) the number of hard negative samples 
$k$, (3) the number of k-means clustering centers $c$, and (4) the contrastive learning loss weight $\alpha_2$.


\textbf{Positive and negative sample distance thresholds.} Table~\ref{hy1} evaluates threshold combinations. The optimal configuration (30 km for positives and 10 km for negatives) yields AUC=0.7436, NDCG@5=0.4345. The 30,km positive threshold captures diverse interactions within a reasonable range. A 10,km negative threshold balances local and distant negatives, avoiding 5,km limitations and 30,km noise. Removing constraints entirely drops the AUC to 0.6811, confirming geographic proximity is critical for sampling.

\textbf{The number of hard negative samples.} Figure~\ref{hy2} (left) evaluates the number of hard negative samples, $k$. Optimal performance occurs at $k=9$ (MRR=0.4266, NDCG@10=0.4996). Smaller values ($k=3,6$; AUC=0.7335, 0.7354) lack enough challenging cases to learn discriminative features. Conversely, larger values ($k=12,15$; AUC=0.7305, 0.7334) introduce noise and overemphasize hard cases, disrupting the learning of general user-item interaction patterns.

\textbf{Number of clustering centers.} Figure~\ref{hy2} (middle) shows the effect of the number of k-means clustering centers ($c \in [5,100]$). A small number of centers (e.g., $c \leq 25$) yields inferior performance, as it cannot adequately capture fine-grained user preferences. Performance improves as $c$ increases and peaks at $c = 75$ with an AUC of 0.7366. However, further increasing $c$ degrades performance (AUC = 0.7330 at $c = 100$), likely because overly fine-grained clustering fragments item semantics and introduces noise.

\textbf{Contrastive loss weight.} Figure~\ref{hy2} (right) shows the impact of the contrastive loss weight $\alpha_2$ in Eq.~\eqref{eq14}. With $\alpha_2 = 1 \times 10^{-5}$, the AUC is 0.7381, indicating limited contribution from the contrastive objective. Increasing $\alpha_2$ to $1 \times 10^{-4}$ and $1 \times 10^{-3}$ improves the AUC to 0.7386 and 0.7399, respectively, proving moderate contrastive supervision enhances representation discrimination. In contrast, setting $\alpha_2$ to $1 \times 10^{-1}$ causes the AUC to decline significantly to 0.7279, with a slight recovery to 0.7357 at $\alpha_2 = 1$. 

\begin{table}[t]
\centering
\small
\caption{Online performance comparison.}
\label{online}
\vspace{-9pt}
\renewcommand{\arraystretch}{1.07}
\setlength{\tabcolsep}{6pt}
\begin{tabular}{l cc cc}
\toprule
\multirow{2}{*}{\textbf{Metrics}} & \multicolumn{2}{c}{\textbf{Trade}} & \multicolumn{2}{c}{\textbf{Experience}} \\
\cmidrule(lr){2-3} \cmidrule(lr){4-5}
 & \textbf{GMV} & \textbf{Orders} & \textbf{Play Time} & \textbf{Clicks} \\
\midrule
\textbf{Ours vs Base} & +2.804\% & +1.285\% & +4.188\% & +1.226\% \\
\bottomrule
\end{tabular}
\vspace{-16pt}
\end{table}

\subsection{Online A/B Test}
We deployed ReST on Kuaishou's local lifestyle short-video platform for online A/B testing (Mar 14–26, 2025; millions of DAUs) to address long-tail content distribution challenges. By integrating ReST into the baseline ranking model, we enhanced ID representations and improved the distribution efficiency of mid- and small-scale items, ultimately boosting overall GMV and total orders.

\textbf{Overall Performance}. 
Table~\ref{online} reports significant transaction and engagement uplifts. ReST increased GMV by \textbf{2.804\%} and Orders by \textbf{1.285\%}. Simultaneously, user experience improved (\textbf{Play Time +4.188\%}, Clicks +1.226\%). This surge indicates ReST effectively captures user preferences, indicating consistent practical gains.


\textbf{Long-tail Performance.} Figure~\ref{vis1} shows that our method improves both orders and GMV across all categories. It brings consistent gains for popular items, with orders increasing by 0.3\%--39.8\% and GMV by 8.7\%--48.2\%. More importantly, it yields substantial improvements for long-tail items, where orders rise by 100.0\%--125.0\% and GMV by 79.6\%--1452.4\%. 
Thus, the method balances exposure and conversions, benefiting less popular categories while comprehensively improving overall transaction performance.

\section{Conclusions}
To address spatial constraints and long-tail effects in local-life recommendation, we propose ReST, a plug-and-play ID representation enhancement framework. Using self-supervised learning tailored to local lifestyle services, ReST overcomes the limitations of traditional methods. Experiments show ReST outperforms state-of-the-art spatiotemporal and long-tail models, with all modules contributing to these gains. Evaluations on cold-start datasets confirm its robustness under sparse interactions and its ability to enhance long-tail item representation. Online A/B testing in a real-world deployment demonstrates significant increases in GMV and order volume, alongside better long-tail recommendation performance. As a plug-and-play module, ReST integrates seamlessly with various backbone models, consistently boosting their performance.


\bibliographystyle{ACM-Reference-Format}
\bibliography{Main.bib}

@String{Springer = "Springer-Verlag" }

@article{oord2018representation,
  title={Representation learning with contrastive predictive coding},
  author={Oord, Aaron van den and Li, Yazhe and Vinyals, Oriol},
  journal={arXiv preprint arXiv:1807.03748},
  year={2018}
}

@article{zhang2025video,
  title={Video corpus moment retrieval with query-specific context learning and progressive localization},
  author={Zhang, Long and Song, Peipei and Duan, Zhangling and Wang, Shuo and Chang, Xiaojun and Yang, Xun},
  journal={IEEE Transactions on Circuits and Systems for Video Technology},
  year={2025},
  publisher={IEEE}
}

@article{song2025towards,
  title={Towards Efficient Partially Relevant Video Retrieval with Active Moment Discovering},
  author={Song, Peipei and Zhang, Long and Lan, Long and Chen, Weidong and Guo, Dan and Yang, Xun and Wang, Meng},
  journal={arXiv preprint arXiv:2504.10920},
  year={2025}
}

@article{zhang2025enhancing,
  title={Enhancing partially relevant video retrieval with robust alignment learning},
  author={Zhang, Long and Song, Peipei and Dong, Jianfeng and Li, Kun and Yang, Xun},
  journal={arXiv preprint arXiv:2509.01383},
  year={2025}
}

@article{wang2025episodic,
  title={Episodic memory representation for long-form video understanding},
  author={Wang, Yun and Zhang, Long and Liu, Jingren and Yan, Jiaqi and Zhang, Zhanjie and Zheng, Jiahao and Yang, Xun and Wu, Dapeng and Chen, Xiangyu and Li, Xuelong},
  journal={arXiv preprint arXiv:2508.09486},
  year={2025}
}

@inproceedings{luo2023improving,
  title={Improving long-tail item recommendation with graph augmentation},
  author={Luo, Sichun and Ma, Chen and Xiao, Yuanzhang and Song, Linqi},
  booktitle={Proceedings of the 32nd ACM international conference on information and knowledge management},
  pages={1707--1716},
  year={2023}
}

@inproceedings{zhang2023empowering,
  title={Empowering long-tail item recommendation through cross decoupling network (CDN)},
  author={Zhang, Yin and Wang, Ruoxi and Cheng, Derek Zhiyuan and Yao, Tiansheng and Yi, Xinyang and Hong, Lichan and Caverlee, James and Chi, Ed H},
  booktitle={Proceedings of the 29th ACM SIGKDD Conference on Knowledge Discovery and Data Mining},
  pages={5608--5617},
  year={2023}
}

@article{yan2025teleego,
  title={TeleEgo: Benchmarking Egocentric AI Assistants in the Wild},
  author={Yan, Jiaqi and Ren, Ruilong and Liu, Jingren and Xu, Shuning and Wang, Ling and Wang, Yiheng and Zhong, Xinlin and Wang, Yun and Zhang, Long and Chen, Xiangyu and others},
  journal={arXiv preprint arXiv:2510.23981},
  year={2025}
}

@article{hidasi2015session,
  title={Session-based recommendations with recurrent neural networks},
  author={Hidasi, Bal{\'a}zs and Karatzoglou, Alexandros and Baltrunas, Linas and Tikk, Domonkos},
  journal={arXiv preprint arXiv:1511.06939},
  year={2015}
}

@article{zhang2022enhancing,
  title={Enhancing sequential recommendation with graph contrastive learning},
  author={Zhang, Yixin and Liu, Yong and Xu, Yonghui and Xiong, Hao and Lei, Chenyi and He, Wei and Cui, Lizhen and Miao, Chunyan},
  journal={arXiv preprint arXiv:2205.14837},
  year={2022}
}

@article{chen2022building,
  title={Building and exploiting spatial--temporal knowledge graph for next POI recommendation},
  author={Chen, Wei and Wan, Huaiyu and Guo, Shengnan and Huang, Haoyu and Zheng, Shaojie and Li, Jiamu and Lin, Shuohao and Lin, Youfang},
  journal={Knowledge-Based Systems},
  volume={258},
  pages={109951},
  year={2022},
  publisher={Elsevier}
}

@inproceedings{shin2024attentive,
  title={An attentive inductive bias for sequential recommendation beyond the self-attention},
  author={Shin, Yehjin and Choi, Jeongwhan and Wi, Hyowon and Park, Noseong},
  booktitle={Proceedings of the AAAI Conference on Artificial Intelligence},
  volume={38},
  number={8},
  pages={8984--8992},
  year={2024}
}

@article{lin2022spatiotemporal,
  title={Spatiotemporal-enhanced network for click-through rate prediction in location-based services},
  author={Lin, Shaochuan and Yu, Yicong and Ji, Xiyu and Zhou, Taotao and He, Hengxu and Sang, Zisen and Jia, Jia and Cao, Guodong and Hu, Ning},
  journal={arXiv preprint arXiv:2209.09427},
  year={2022}
}

@article{schafer2001commerce,
  title={E-commerce recommendation applications},
  author={Schafer, J Ben and Konstan, Joseph A and Riedl, John},
  journal={Data mining and knowledge discovery},
  volume={5},
  pages={115--153},
  year={2001},
  publisher={Springer}
}

@incollection{celma2010music,
  title={Music recommendation},
  author={Celma, Oscar},
  booktitle={Music recommendation and discovery: The long tail, long fail, and long play in the digital music space},
  pages={43--85},
  year={2010},
  publisher={Springer}
}

@misc{hu2019squeezeandexcitationnetworks,
      title={Squeeze-and-Excitation Networks}, 
      author={Jie Hu and Li Shen and Samuel Albanie and Gang Sun and Enhua Wu},
      year={2019},
      eprint={1709.01507},
      archivePrefix={arXiv},
      primaryClass={cs.CV},
      url={https://arxiv.org/abs/1709.01507}, 
}

@inproceedings{trisan,
  title={Trilateral spatiotemporal attention network for user behavior modeling in location-based search},
  author={Qi, Yi and Hu, Ke and Zhang, Bo and Cheng, Jia and Lei, Jun},
  booktitle={Proceedings of the 30th ACM International Conference on Information \& Knowledge Management},
  pages={3373--3377},
  year={2021}
}

@inproceedings{ofrs,
  title={Exploring the spatiotemporal features of online food recommendation service},
  author={Lin, Shaochuan and Pei, Jiayan and Zhou, Taotao and He, Hengxu and Jia, Jia and Hu, Ning},
  booktitle={Proceedings of the 46th International ACM SIGIR Conference on Research and Development in Information Retrieval},
  pages={3354--3358},
  year={2023}
}

@inproceedings{basm,
  title={BASM: A Bottom-up Adaptive Spatiotemporal Model for Online Food Ordering Service},
  author={Du, Boya and Lin, Shaochuan and Gao, Jiong and Ji, Xiyu and Wang, Mengya and Zhou, Taotao and He, Hengxu and Jia, Jia and Hu, Ning},
  booktitle={2023 IEEE 39th International Conference on Data Engineering (ICDE)},
  pages={3549--3562},
  year={2023},
  organization={IEEE}
}

@article{dropoutnet,
  title={Dropoutnet: Addressing cold start in recommender systems},
  author={Volkovs, Maksims and Yu, Guangwei and Poutanen, Tomi},
  journal={Advances in neural information processing systems},
  volume={30},
  year={2017}
}

@inproceedings{sigir19,
  title={Warm up cold-start advertisements: Improving ctr predictions via learning to learn id embeddings},
  author={Pan, Feiyang and Li, Shuokai and Ao, Xiang and Tang, Pingzhong and He, Qing},
  booktitle={Proceedings of the 42nd International ACM SIGIR Conference on Research and Development in Information Retrieval},
  pages={695--704},
  year={2019}
}

@inproceedings{cl4srec,
  title={Contrastive learning for sequential recommendation},
  author={Xie, Xu and Sun, Fei and Liu, Zhaoyang and Wu, Shiwen and Gao, Jinyang and Zhang, Jiandong and Ding, Bolin and Cui, Bin},
  booktitle={2022 IEEE 38th international conference on data engineering (ICDE)},
  pages={1259--1273},
  year={2022},
  organization={IEEE}
}

@inproceedings{lot-crs,
  title={Alleviating the long-tail problem in conversational recommender systems},
  author={Zhao, Zhipeng and Zhou, Kun and Wang, Xiaolei and Zhao, Wayne Xin and Pan, Fan and Cao, Zhao and Wen, Ji-Rong},
  booktitle={Proceedings of the 17th ACM Conference on Recommender Systems},
  pages={374--385},
  year={2023}
}

@inproceedings{him,
  title={Hierarchical Interest Modeling of Long-tailed Users for Click-Through Rate Prediction},
  author={Xie, Xu and Niu, Jin and Deng, Lifang and Wang, Dan and Zhang, Jiandong and Wu, Zhihua and Bian, Kaigui and Cao, Gang and Cui, Bin},
  booktitle={2023 IEEE 39th International Conference on Data Engineering (ICDE)},
  pages={3058--3071},
  year={2023},
  organization={IEEE}
}

@inproceedings{din,
  title={Deep interest network for click-through rate prediction},
  author={Zhou, Guorui and Zhu, Xiaoqiang and Song, Chenru and Fan, Ying and Zhu, Han and Ma, Xiao and Yan, Yanghui and Jin, Junqi and Li, Han and Gai, Kun},
  booktitle={Proceedings of the 24th ACM SIGKDD international conference on knowledge discovery \& data mining},
  pages={1059--1068},
  year={2018}
}

@inproceedings{rcl,
  title={Relative Contrastive Learning for Sequential Recommendation with Similarity-based Positive Sample Selection},
  author={Wang, Zhikai and Shen, Yanyan and Zhang, Zexi and He, Li and Li, Yichun and Gu, Hao and Zhang, Yinghua},
  booktitle={Proceedings of the 33rd ACM International Conference on Information and Knowledge Management},
  pages={2493--2502},
  year={2024}
}

@inproceedings{elemedataset,
  title={Spatial-Temporal Knowledge Distillation for Takeaway Recommendation},
  author={Zhao, Shuyuan and Chen, Wei and Shi, Boyan and Zhou, Liyong and Lin, Shuohao and Wan, Huaiyu},
  booktitle={Proceedings of the AAAI Conference on Artificial Intelligence},
  volume={39},
  number={12},
  pages={13365--13373},
  year={2025}
}

@article{auc,
  title={The meaning and use of the area under a receiver operating characteristic (ROC) curve.},
  author={Hanley, James A and McNeil, Barbara J},
  journal={Radiology},
  volume={143},
  number={1},
  pages={29--36},
  year={1982}
}

@inproceedings{mrr,
  title={The trec-8 question answering track report.},
  author={Voorhees, Ellen M and others},
  booktitle={Trec},
  volume={99},
  pages={77--82},
  year={1999}
}

@inproceedings{he2017neural,
  title={Neural collaborative filtering},
  author={He, Xiangnan and Liao, Lizi and Zhang, Hanwang and Nie, Liqiang and Hu, Xia and Chua, Tat-Seng},
  booktitle={Proceedings of the 26th international conference on world wide web},
  pages={173--182},
  year={2017}
}

@article{luo2024qarm,
  title={QARM: Quantitative Alignment Multi-Modal Recommendation at Kuaishou},
  author={Luo, Xinchen and Cao, Jiangxia and Sun, Tianyu and Yu, Jinkai and Huang, Rui and Yuan, Wei and Lin, Hezheng and Zheng, Yichen and Wang, Shiyao and Hu, Qigen and others},
  journal={arXiv preprint arXiv:2411.11739},
  year={2024}
}

@article{robusto1957cosine,
  title={The cosine-haversine formula},
  author={Robusto, C Carl},
  journal={The American Mathematical Monthly},
  volume={64},
  number={1},
  pages={38--40},
  year={1957},
  publisher={JSTOR}
}

@article{wang2025fim,
  title={FIM: Frequency-Aware Multi-View Interest Modeling for Local-Life Service Recommendation},
  author={Wang, Guoquan and Luo, Qiang and Hu, Weisong and Yao, Pengfei and Zeng, Wencong and Zhou, Guorui and Gai, Kun},
  journal={arXiv preprint arXiv:2504.17814},
  year={2025}
}

@article{chi2023modeling,
  title={Modeling spatiotemporal periodicity and collaborative signal for local-life service recommendation},
  author={Chi, Huixuan and Xu, Hao and Liu, Mengya and Bei, Yuanchen and Zhou, Sheng and Liu, Danyang and Zhang, Mengdi},
  journal={arXiv preprint arXiv:2309.12565},
  year={2023}
}

@article{lecun2015deep,
  title={Deep learning},
  author={LeCun, Yann and Bengio, Yoshua and Hinton, Geoffrey},
  journal={nature},
  volume={521},
  number={7553},
  pages={436--444},
  year={2015},
  publisher={Nature Publishing Group UK London}
}

@inproceedings{liu2014geohash,
  title={A geohash-based index for spatial data management in distributed memory},
  author={Liu, Jiajun and Li, Haoran and Gao, Yong and Yu, Hao and Jiang, Dan},
  booktitle={2014 22Nd international conference on geoinformatics},
  pages={1--4},
  year={2014},
  organization={IEEE}
}

@techreport{arthur2006k,
  title={k-means++: The advantages of careful seeding},
  author={Arthur, David and Vassilvitskii, Sergei},
  year={2006},
  institution={Stanford}
}

\section{Main author bio}

\textbf{Hao Jiang} is currently a Researcher at Kuaishou Technology, focusing on applying Large Language Models to short-video ranking and recommendation scenarios. He received both his B.E. and M.E. degrees from the School of Information and Communication Engineering at the Communication University of China (CUC). 

\end{document}